\documentclass[sigconf]{acmart}
\AtBeginDocument{%
  }

\usepackage{tabularx} 
\usepackage{array}
\begin{document}

\title{Linguistically Augmented Audio Speech Data (LinguAS)}

\author{Ashley R. Keaton}
\email{arkeaton@umbc.edu}
\affiliation{%
  \institution{University of Maryland, 
Baltimore County}
  \city{Baltimore}
  \state{MD}
  \country{USA}
}

\author{Zahra Khanjani}
\email{zkhanja1@umbc.edu}
\affiliation{%
  \institution{University of Maryland, 
Baltimore County}
  \city{Baltimore}
  \state{MD}
  \country{USA}
}

\author{Christine Mallinson}
\email{mallinson@umbc.edu}
\affiliation{%
  \institution{University of Maryland, 
Baltimore County}
  \city{Baltimore}
  \state{MD}
  \country{USA}
}

\author{Vandana P. Janeja}
\authornote{Corresponding Author}
\email{vjaneja@umbc.edu}
\affiliation{%
  \institution{University of Maryland, 
Baltimore County}
  \city{Baltimore}
  \state{MD}
  \country{USA}
}

\renewcommand{\shortauthors}{Keaton, Khanjani et al.}

\begin{abstract}
 Maliciously-created fake speech, including deepfaked and spoofed audio, is proliferating at an alarming rate, and detection models are racing to stay ahead of the curve. Yet, most detection models are trained to make inference on frame-level audio features alone without leveraging valuable linguistic cues at larger timescales. To address this gap, we present Linguistically Augmented Audio Speech Data (LinguAS), a dataset of genuine and deepfaked audio samples annotated with five strategically-chosen, Expert-Defined Linguistic Features (EDLFs) that occur frequently in spoken English and are characteristic of natural human speech. LinguAS contains over 800 audio samples, each of which are annotated with EDLFs. The dataset has a balanced number of four spoofed audio attack types and a proportionate number of genuine speech samples. We also include metadata on speaker gender and the generator/source for each spoofed audio sample, offering more granularity for model training. We found that models trained on data augmented with EDLFs had improved model performance significantly beyond the ASVspoof 2021 deep learning baselines and SSL models like HuBert and XLSR. LinguAS’s augmented linguistic, gender, and generator metadata provide audio deepfake researchers with a dataset that emphasizes real human language traits to improve model inference of faked speech. Data and code are publicly available at  \cite{figs} and \cite{code}.
\end{abstract}

\keywords{spoofed audio detection, deepfakes, audio deepfakes, linguistics, linguistic features
}


\maketitle

\section{Introduction}
Synthetic speech generation methods have increased in sophistication in recent years, even becoming perceptually indistinguishable from human speech for some listeners \cite{muller}. While these hyperrealistic speech generation models have exciting implications for intelligent voice assistants, clinical voice prosthesis, and education, they also pose serious concerns for the proliferation of audio deepfakes. An audio deepfake is an unauthorized reproduction of a person’s unique voice characteristics and are increasingly used in scams. From 2024 to 2025, the voice identification company Pindrop reported a 1,300\% increase in voice deepfake scams \cite{Pindrop2025}. In light of this alarming increase of spoofed audio generation, we present the Linguistically Augmented Audio Speech Data (LinguAS) dataset. We approach deepfake detection from an interdisciplinary collaboration with expert language scientists to augment benchmark audio speech datasets with linguistically-annotated data that can be leveraged for more sophisticated model inference beyond feature- or frame-level acoustic anomaly detection. This new avenue of dataset annotation strategically tags language features across multiple timescales and can be used to model interactions between frame-level anomalies and larger timescale features, or interactions between linguistic features. LinguAS is an exemplar dataset for a new approach to answering data science questions with small but powerfully-designed datasets that contain detailed, structured information that generalizes robustly to unseen data through transfer learning. The LinguAS dataset is open access \cite{figs} and the benchmarking code \cite{code} is publicly available. The data is available at \cite{dframe}.
The LinguAS dataset makes these main contributions:
\begin{itemize}
    \item Provides a unique feature set of expert-defined linguistic features that are associated with spoofed audio.
    \item Encodes different levels of linguistic representations, spanning from phonetic to prosodic features. 
\item Improves model fine-tuning and explainability by enabling researchers to test which linguistic features are most predictive of spoofed audio.
\item Mitigates the possibility of overfitting to dataset- or generator- specific cues.
\end{itemize}

\section{Background and Motivation}
The LinguAS dataset aims to address the current assumption in scientific modeling that larger datasets are inherently better than smaller ones. As one example, the popular LJspeech dataset contains over 13,000 samples, but all were produced by the same human speaker \cite{Ito2017LJSpeech}. A model trained on LJspeech alone will learn the features of that particular speaker’s voice qualities with a high degree of granularity, but will generalize poorly onto other real speakers’ voices. While it is widely understood that modeling a single person’s speech will result in a bad language model, audio deepfake training datasets frequently include many instances of synthetic audio from a small set of text-to-speech (TTS) or voice conversion (VC) algorithms and appear misleadingly diverse. Hong et al. \cite{Hong2025} found that models trained on high-fidelity spoofed speech learned subtle, generator-specific cues which did not generalize to out-of-distribution data, even for lower-fidelity spoofed audio samples. They argue that currently, too much attention is placed on “the right model” or “the right data” rather than a focus on a wide array of audio sample diversity. 
While the risk of overfitting to speaker- or generator- specific traits has gained attention in recent years, a potentially powerful avenue of generalization remains understudied: abstracting frame-level acoustic features into higher-order linguistic features for inference. Because machine learning deepfake detection models typically make inference on the frame level, they overlook meaningful structural features of language that can improve inference.

The LinguAS data can also be used to address the computational cost constraints on deepfake detection model development. A recent systematic review \cite{Akhtar2024} highlight the importance of preprocessing and hybrid processing steps to improve scalability of detection models. We demonstrate that LinguAS feature annotations can scaffold development of cost-efficient ensemble models with computationally light pre-filters to reduce computational load on heavier models upstream. Finally,  the inclusion of high-level, human-interpretable linguistic features in LinguAS enables researchers to develop more trustworthy models, both in terms of robustness and explainability.  

\subsection{Linguistic representation and knowledge}
In audio sample modeling for deepfake detection, there are two typical approaches to audio preprocessing.  One common approach is to model the raw acoustic audio without any transformation.  More recently, the most common preprocessing technique is to Fourier-transform a raw audio signal into a frame-level representation where acoustic information is represented visually \cite{Sahidullah}. Linear-frequency cepstral coefficients (LFCCs) and Mel-frequency cepstrum coefficients (MFCCs) are both frame-level representations of audio and frequently used for audio deep learning \cite{Khochare2021}.
	The field of linguistics organizes different levels of speech representation hierarchically, from the most basic production of speech sounds (known as acoustic phonetics) to the entire conversational context (known as discourse analysis or pragmatics). In the LinguAS dataset, we include linguistic annotations that span this hierarchy. We include the acoustic phonetic feature of breath, the phonological  feature of stop consonant burst production, the phrase-level features of pause and pitch variation, and audio recording naturalness across the entire speech sample.
	Importantly, acoustic and frame-level representations of speech are complementary to linguistic representations. In the subfield of acoustic phonetics, LFCCs and MFCCs can be the primary source of data, and entire research programs might be based around a linguistic phenomena that is <10ms long \cite{Iskarous2025}. The primary difference between linguistic and machine learning features is how “experts”– linguists and computers, respectively– encode acoustic information into meaningful classifications of language. 
The LinguAS data illustrates how domain expert knowledge can improve model inference by augmenting data with annotations that correspond to real-world knowledge that an algorithm is unlikely to encode. Put simply, linguists dedicate their entire lives to listening to how people talk. When tasked with identifying anomalous features in naturalistic deepfaked speech, linguists have the training to notice subtle acoustic differences and also talk about them in a systematic way.

\subsection{Related work}
Related work has found that the inclusion of linguistic feature annotation can significantly improve model performance. Blue et al. \cite{Blue2022} leveraged the physiological limitations of what a human vocal tract can produce for comparison against acoustic waveforms. By modeling the anatomical plausibility of the transition between phoneme bigrams, Blue et al.\cite{Blue2022} achieved 99.9\% precision in their deepfake detection model. More recently,  Zhang et al. \cite{Zhang2025} found that model inference based on adaptive pooling of phonemes outperformed state-of-the-art benchmark models that used frame-level representations. Other studies have investigated how the correlation between breath, speech and silence can predict whether the sample is real human speech \cite{Doan2023, Mostaani2022}. These examples illustrate that training models on high-level linguistic features can improve model performance on unseen data. 
Another approach to linguistic feature representation in deepfake detection is self-supervised feature embedding developments \cite{MartinDonas2024}. Research on self-supervised speech embeddings also shows promising improvements to model performance, but still face the issues of poor explainability and the need for a large training dataset. The data presented in LinguAS is explainable and can be used to model linguistic features at multiple different timescales, including pitch variation and phoneme segments.
More generally, a “data centric” AI movement that focuses on small, explainable datasets is developing across scientific disciplines \cite{Strickland}. Molecular biology and genetics, disciplines which frequently benefit from artificial intelligence applications, are often bound to small datasets because of cost, lack of data availability, or privacy constraints. 

We argue that the number of observations in a dataset is not proportionate to the amount of information it contributes to answering a research question. Instead, the informative power of a dataset lies in how “deep” the data is– beyond the number of samples, how much meaningful information does the dataset contain? To test this question, we developed the LinguAS dataset. In the following section we describe how we curated LinguAS, a compact but information-dense dataset for deepfake detection modeling. 

\section{LinguAS Data Creation Approach}
The main aim of the LinguAS dataset is to curate a collection of diverse audio samples of spoofed and bona fide speech, and find generalizable patterns of spoofed audio cues across a wide variety of speech samples. We approached this research goal in two ways: 
\begin{enumerate}
    \item Dataset breadth: Create a dataset of audio samples balanced across fake and genuine audio, different spoof attack type, and speaker gender, and 
\item Dataset depth: Quantify expert linguists’ perceptions of anomalous audio cues into an operationalizable format for machine and deep learning models. 

\end{enumerate}
The data is publicly available at \cite{figs}. 

\subsection{Dataset breadth}
As mentioned in section 2, dataset diversity is essential to accurate out-of-sample prediction of spoofed audio.  We curated sources, speakers, and attack types so that the LinguAS dataset contains a wide array of human (and human-emulating) speech variation. This is especially important given the relatively small size of the LinguAS dataset. We curated our audio samples by proportion of spoofed versus genuine audio, attack type (voice conversion, text to speech, replay attack, and mimicry), and speaker gender (shown in Appendix A.1 Table \ref{tab:appendix_1}).
 The representation of different attack types within one dataset is a unique benefit of the LinguAS dataset, as many deepfake datasets contain only one or a limited set of attack types \cite{Reimao2019FoR, Yamagishi2021ASVspoof2021}. In particular, the LinguAS dataset contains samples of both artificial intelligence (AI)-generated speech and human-produced manipulated speech. Replay attacks and mimicry are recordings of real human speech, yet are often treated in detection models as “fake” speech alongside TTS and VC. Thus, cues for deepfaked versus replay/mimic spoofs have qualitatively different markers of voice spoofing, which may account for the known differences in model performance on replay attacks compared to deepfakes \cite{Huang2019, Lai2019, Kinnunen2017ASVspoof}. Finally, to mitigate against speaker- or generator-specific cues in detection models, we curated the dataset to contain a wide variety of both human voices and deepfake generators.
 
\subsubsection{Audio sample curation } The LinguAS dataset combines audio samples from publicly-available benchmark datasets with audio samples collected by the LinguAS research team. We created samples from several different TTS and VC generators to diversify the TTS and VC samples already present in FoR \cite{Reimao2019FoR} and ASVspoof 2021 \cite{Yamagishi2021ASVspoof2021}. We also collected Youtube audio of public figures and their impersonators so that we could directly compare genuine speech to mimicry attacks. Table {\ref{tab_source_of_audio}} lists the audio sample sources we used to curate the LinguAS dataset. 


\begin{table*}[]
\caption{Source of Audio Samples in LinguAS Dataset}
\label{tab_source_of_audio}
\begin{tabular}{|l|l|}
\hline
\textbf{Publicly-Available datasets}   & \textbf{Generated/collected by LinguAS team} \\ \hline
LJSpeech (Ito and Johnson \cite{Ito2017LJSpeech}) & ASSEM-VC (Kim et al.\cite{Kim2022AssemVC})                              \\ \hline
ASVspoof 2021 (Yamagishi et al.\cite{Yamagishi2021ASVspoof2021}) & MelGan (Kumar et al. \cite{Kumar2019Melgan}                 \\ \hline
ASVspoof 2017 (Wu et al.) \cite{Wu2017ASVspoof} & Cotatron (Park et al.)   \cite{Park2020Cotatron}                          \\ \hline
FakeOrReal (Reimao and  Tzerpos ) \cite{Reimao2019FoR} & Mellotron (Valle et al.)  \cite{Valle2020Mellotron}             \\ \hline
                                 & Wavenet (Van den Oord et al. )  \cite{VanDenOord2016Wavenet}                   \\ \hline
                                 & ResembleAI                                               \\ \hline
                                 & Google TTS                                               \\ \hline
                                 & Youtube videos of public figures and their impersonators \\ \hline
\end{tabular}
\end{table*}

\subsection{Dataset depth}
\subsubsection{Strategic linguistic labeling.} Beyond the diversity of audio samples, the key innovation of the LinguAS dataset is the augmentation of spoofed and real audio samples with linguistic annotation data. We decided a priori to create five linguistic features that each audio sample would be labeled with, and that we wanted to include linguistic representations at different timescales. To determine which features to include, three linguists on the team independently listened to audio clips and qualitatively assessed what themes appeared in their perceptions of spoofed audio. Based on their shared perceptions, they focused their analysis on the five features defined in Table {\ref{tab_typical_expected_duration}}. 

\begin{table*}[]
\caption{Linguistic representations of each of the EDLFs and their typical expected duration in natural human speech}
\label{tab_typical_expected_duration}
\begin{tabular}{|l|l|l|}
\hline
\multicolumn{1}{|c|}{\textbf{EDLF}} & \multicolumn{1}{c|}{\textbf{Linguistic Representation Level}} & \multicolumn{1}{c|}{\textbf{Approx. Typical Timescale}} \\ \hline
Intake/outtake of breath    & Phonetic                       & 1–2 seconds, variable to context \\ \hline
Stop consonant bursts       & Phonological                   & $\sim$10ms                       \\ \hline
Pitch variation             & Lexical, Phrasal or Sentential & Variable to context              \\ \hline
Pause                       & Phrasal                        & Variable to context              \\ \hline
Audio recording naturalness & Global                         & Entire length of sample          \\ \hline
\end{tabular}
\end{table*}

The LinguAS dataset leverages domain expert knowledge from linguistics that speakers vary their speech in predictable ways. For example, phonological studies have extensively shown that the voicing distinction in the English stop consonant phonological class $(/p/ and /b/; /t/ and /d/; /k/ and /g/)$ is primarily differentiated by high-frequency spectral energy \cite{Chodroff2014}. Importantly, naturally-produced speech errors are also predictable \cite{WestburyKeating1986}. Thus, they hypothesized that naturally-produced human speech errors are reliably qualitatively different from artifacts in spoofed audio samples. 
 A widely understood property of speech in linguistics is the relationship between a person’s natural pitch, known as fundamental frequency (f0), and other phonetic characteristics of their voice. Crucially, a person’s f0 impacts how they produce many other phonetic traits, which makes interactions between f0 and other linguistic features a good opportunity to listen for sources of speech alteration. When a speaker makes an intentional effort to raise or lower the pitch of their voice, this is described as pitch variation.  Pitch variation anomaly was defined as an EDLF because TTS voices have characteristically flatter pitch than human speakers, and transient pitch excursions can appear during VC.
The breath anomaly EDLF was chosen because breath is multivariate with many other linguistic productions. Breath sounds should appear at typical time intervals and are also phonetically related to a person’s f0. But, breath length also varies proportionally to anticipated sentence length and also contextual factors like socially cuing the intention to speak next \cite{Kallay2019, RochetCapellan2014}. 
The expert linguists chose audio quality as one of the five features based on their qualitative assessment that the most frequent indicator of manipulated speech was the presence of acoustic artifacts that surfaced when the audio was spoofed. In their qualitative assessments, the linguists consistently mentioned that spoofed samples sounded compressed, muffled, tinny, echoic, or “squashed,” or contained transient acoustic artifacts like splicing clicks. The summary of linguistic criteria for selecting a feature is summarized in {\ref{tab_edlfs_summsry}}.

\subsubsection{Expert Linguist Annotation}
	Three linguists on the team (one professor of linguistics and two graduate students) independently annotated each audio clip. EDLFs were evaluated using a binary score, where 0 is not anomalous and 1 is anomalous. For each audio sample, the linguist would independently listen to the audio clip and evaluate each of the five EDLFs with binary scoring. Any perceptible presence of an anomaly resulted in a 1 score on that feature. After the EDLF binary scoring was complete, the linguists met to assess their level of agreement across feature annotations. Out of the 800 samples in the dataset, there were only four instances of disagreement within the linguists' evaluations. When the linguists had different anomaly perceptions for a linguistic feature, they discussed which label was most appropriate for that sample to reach a unanimous conclusion.

\begin{table*}[t]
\caption{Expert-Defined Linguistic Feature criteria}
\centering
\label{tab_edlfs_summsry}
\begin{tabularx}{\textwidth}{|l|X|}
\hline
\textbf{Expert-Defined Linguistic Feature} &
\textbf{Linguistic Criteria} \\ \hline

Burst anomaly &
The speaker’s productions of stop consonants were unexpected or contextually inappropriate for that speech sound.* \\ \hline

Pause anomaly &
Contextually-inappropriate or unexpectedly short or long pauses, including across word, phrase, and sentence boundaries. \\ \hline

Breath anomaly &
Intake or outtake of breath not typical of human speech, including breaths acoustically mismatched from the speaker’s voice, inconsistent breath rhythm, or an implausibly long speech segment without breath sounds. \\ \hline

Pitch anomaly &
Unexpected or contextually-inappropriate pitch excursion within or across words. Or, a global absence of pitch variation atypical of human speech. \\ \hline

Audio quality anomaly &
Audio that was compressed, tinny, echoic, robotic, oversmoothed, or unnaturally nasal. \\ \hline

\multicolumn{2}{|l|}{\footnotesize *See above for description of stop consonant burst voicing in English.} \\ \hline
\end{tabularx}
\end{table*}

The EDLF evaluation criteria are impressionistic by design. Existing work modeling real and fake speech has shown reliable acoustic differences across voice types. Meanwhile, linguists have a systematic understanding of the underlying reason that a deepfaked voice “doesn’t sound right.” The EDLF coding system links perceptual knowledge from linguists to acoustic cues for deepfaked speech. 
In the next section, we show that EDLFs are successful at linking linguists’ perceptual knowledge to acoustic phenomena. First, we show that machine learning models that use EDLFs are more successful at voice spoofing detection than the ASVspoof 2021 baseline acoustic models. Next, we illustrate that the baseline acoustic models’ performance improves when they are combined into an ensemble model featuring an EDLF classifier. Finally, we show that the performance improvement of EDLFs extends even to self-supervised learning models, illustrating that EDLF classification is generalizable to unseen data and also beneficial to a variety of modeling methods.  

\section{ Benchmark and Validation Experiments}
We validated the benefit of EDLF augmentation in the LinguAS dataset through a series of experiments. For every experiment, the model was trained on  731 samples (87\% of the data) and tested on 134 unseen samples.  For each model, we used k-fold cross validation (k =5) to tune the hyperparameters. 

\subsection{Expert-Defined Linguistic Feature (EDLF) Validation}
\begin{figure}
 \includegraphics[width=.5\textwidth]{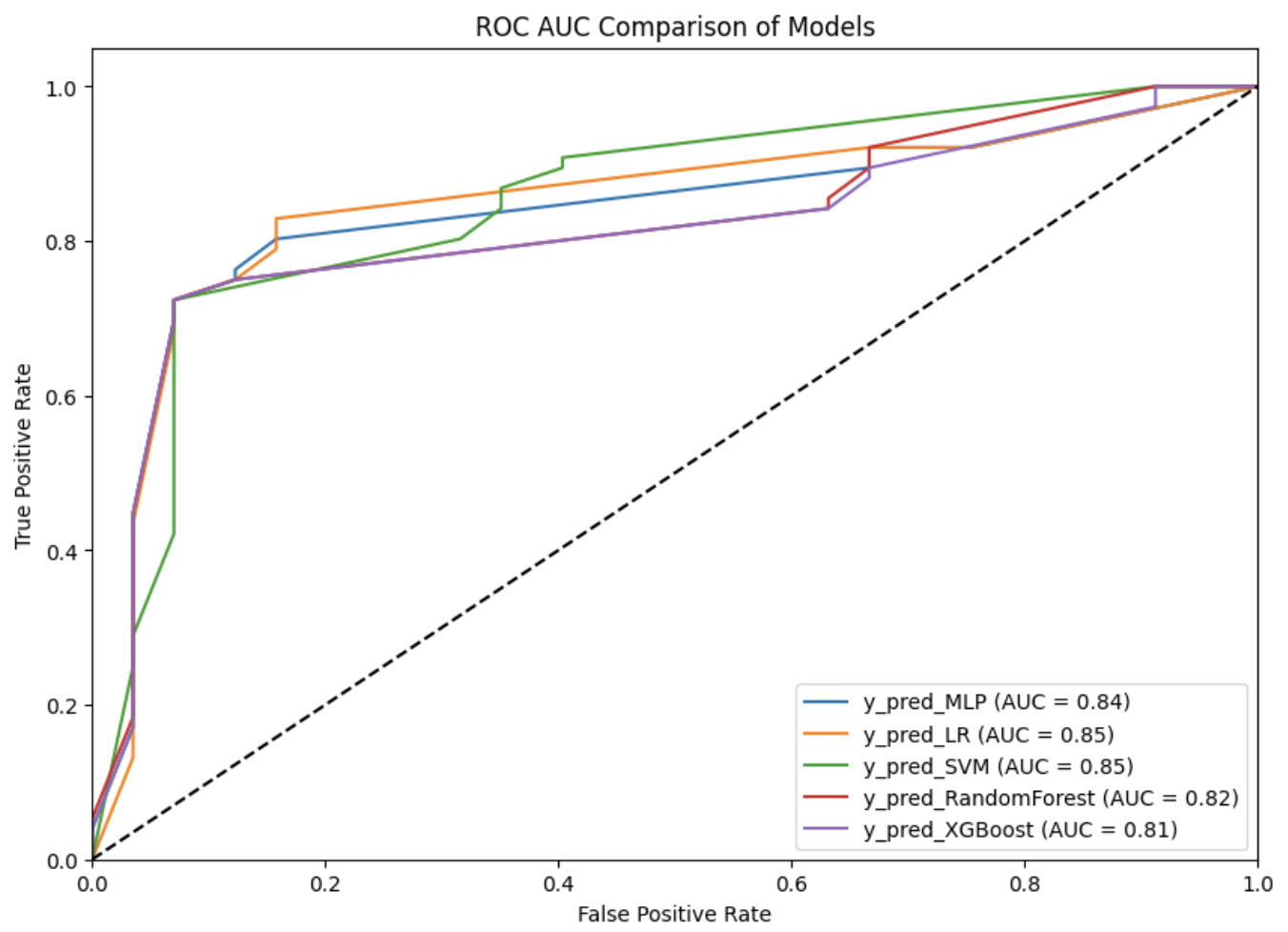}
  \caption{Expert-Defined Linguistic Feature (EDLF) Validation}
  \label{mlmodels}
\end{figure}

A basic empirical question is whether the EDLFs we proposed actually result in model performance benefits. To test this basic premise, we modeled the expert-defined linguistic features (EDLFs) using five traditional machine learning (ML) methods: linear regression, multilayer perceptron, support vector machine, random forest, and  XGBoost. As shown in {\bf Figure \ref{mlmodels}}, we found that the two best-performing models were logistic regression (LR) and support vector machine (SVM), both with AUC scores of 0.85. These findings are consistent with our previous work that found logistic regression (LR) was the best-performing ML method for EDLF-augmented data \cite{Khanjani2023}. 

\subsubsection{Comparison with  ASVspoof baselines}
\begin{figure}
 \includegraphics[width=.5\textwidth]{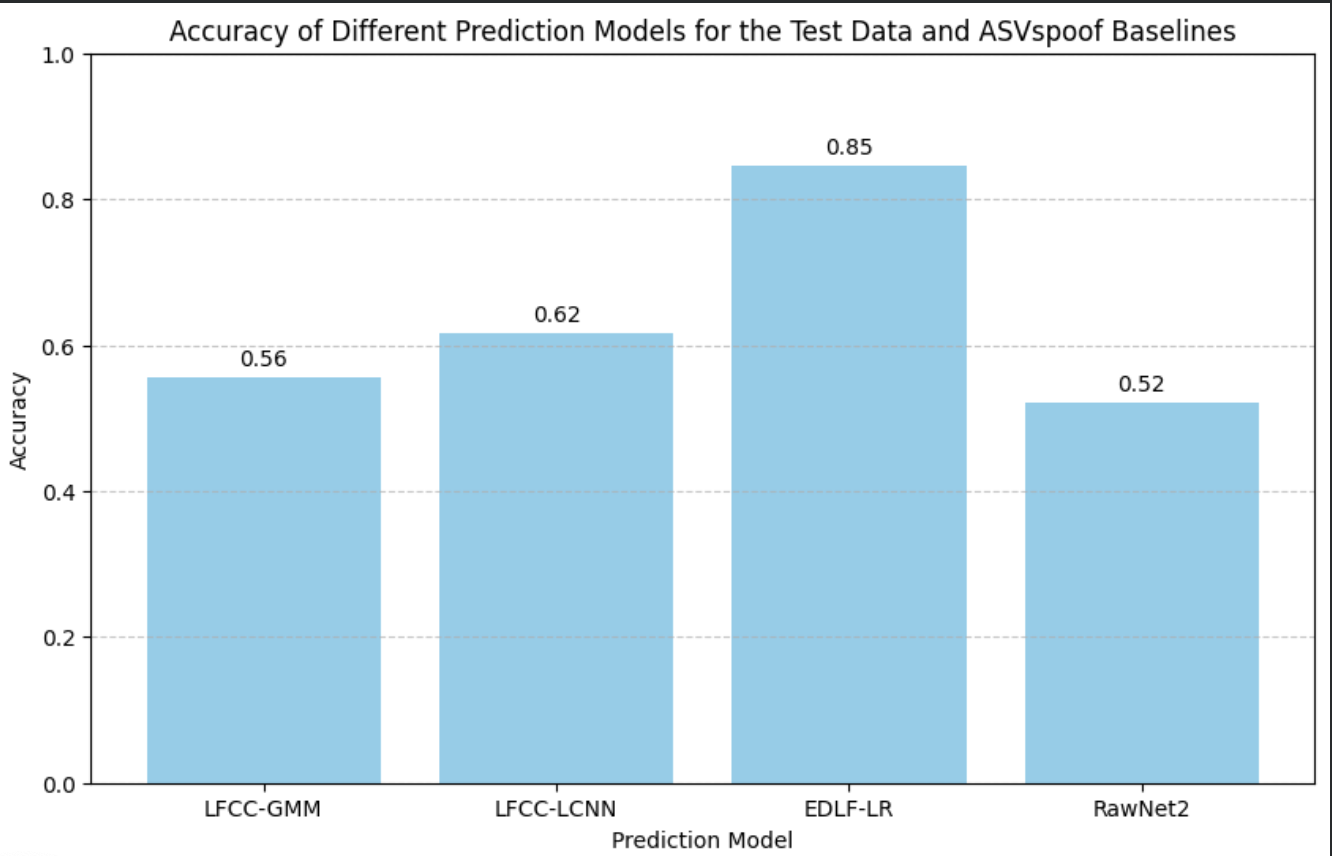}
  \caption{Comparison to ASVspoof baselines}
   \label{accbase}
\end{figure}

Next, we tested our claim that linguistic features provide a more robust benefit to detection models compared to frame-level or acoustic feature modeling. We compared the EDLF-LR model to the ASVSpoof 2021 baseline models which are acoustic- and frame-level deep learning models. The baseline models include an LFCC Gaussian mixture model (GMM), an LFCC light convolutional neural network (LCNN), and RawNet2 \cite{Wang2021Comparative}. {\bf Figure \ref{accbase}}  illustrates the performance improvement of the EDLF-LR model over the ASVspoof baseline models. 

\begin{figure}
 \includegraphics[width=.5\textwidth]{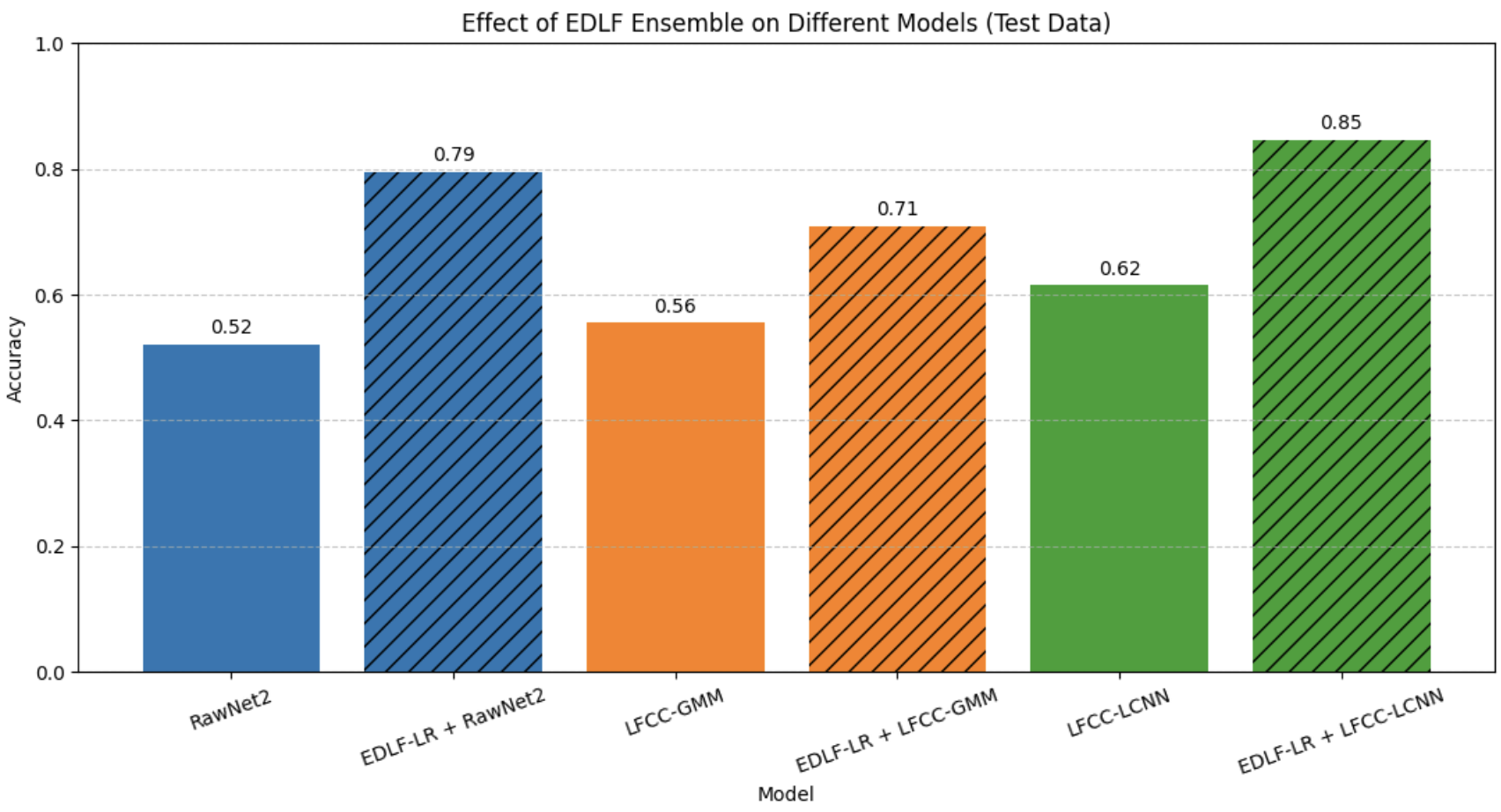}
  \caption{EDLF Ensembles}
   \label{edlfens}
\end{figure}

The  performance benefit of the EDLF-LR model over the acoustic and deep learning models is attractive, but a limitation of ML models is the need for careful feature classification and preprocessing.  Almutairi et al.  \cite{Almutairi2022} suggest that deep learning models are more appropriate for deepfake detection tasks because of their generalizability. With this in mind, we tested whether EDLF-LR classification can be integrated into deep learning models and improve DL model performance. We combined the EDLF-LR model with each of the three ASVspoof baseline models (LFCC-LCNN, LFCC-GMM, and RawNet2). Ensemble models that combined the baseline model with EDLF-LR resulted in increased prediction accuracy for every model we tested, as shown in {\bf Figure \ref{edlfens}}.

\subsection{Performance improvements for deep learning models}
After finding performance improvements for each of the ASVspoof baseline models, we asked if the EDLF-LR augmentation step would also improve performance of SOTA deep learning models ensembled with the EDLF LR. We chose the deep learning model VGGish and two self-supervised learning models, Hubert and XSLR.
We found that EDLF-LR had greater accuracy and specificity than each of the  models we tested. 

\begin{figure}
 \includegraphics[width=.5\textwidth]{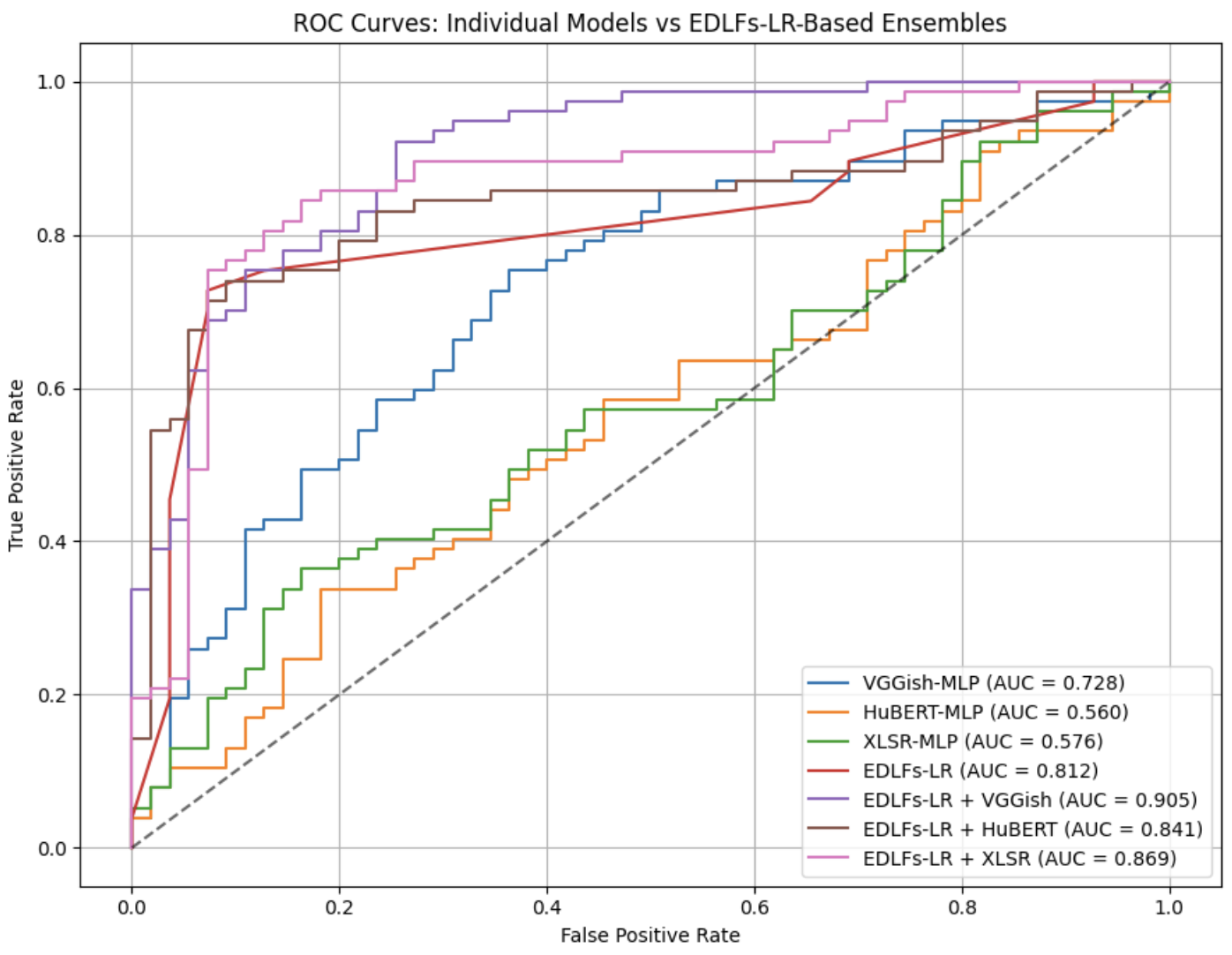}
  \caption{SSL models compared to ELDF-LR}
   \label{dl}
\end{figure}

\begin{figure}
 \includegraphics[width=.5\textwidth]{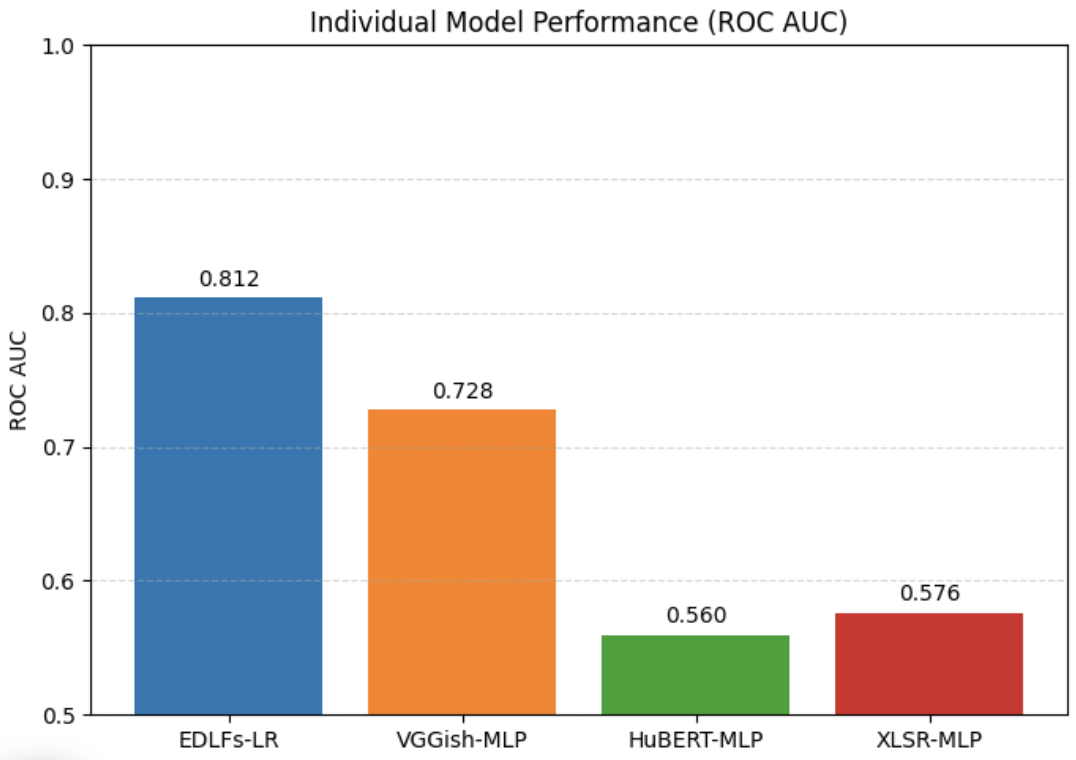}
  \caption{ROC-AUC comparison for SSL based models, SSL+EDLF-LR ensemble models, and EDLF-LR classification}
   \label{dlauc}
\end{figure}

{\bf Figure \ref{dl}} shows the AUC values compared for EDLF-LR and each SOTA model. We then created ensemble models with  EDLF-LR for each of the SOTA models. {\bf Figure \ref{dlauc}} plots the ROC-AUC scores for each SOTA model, SOTA + EDLF-LR ensemble models, and EDLF-LR. Similar to our findings for the ASVspoof baseline models, ensemble methods improved the prediction accuracy of the SOTA models as well. Here, we find that VGGish + EDLF-LR and XLSR + EDLF-LR achieve even better performance than EDLF-LR alone. This suggests that SSL models trained with EDLF-augmented data can generalize the underlying linguistic knowledge represented by the EDLF coding onto unseen data. 
One of the widely adopted evaluation metrics for audio deepfake detection systems is the Equal Error Rate (EER). Table~\ref{tab:eer_dl_ssl} presents the EER comparison among deep learning (DL) and self-supervised learning (SSL) baseline models.

\begin{table}[ht]
\centering
\caption{EER Comparison of DL and SSL Baselines}
\label{tab:eer_dl_ssl}
\begin{tabular}{lccc}
\hline
\textbf{Model} & \textbf{EER} & \textbf{Threshold} & \textbf{ROC AUC} \\
\hline
EDLFs-LR    & 0.1870 & 0.3080 & 0.8119 \\
VGGish-MLP  & 0.3325 & 0.5407 & 0.7275 \\
XLSR-MLP    & 0.4325 & 0.5788 & 0.5764 \\
HuBERT-MLP  & 0.4610 & 0.5241 & 0.5596 \\
\hline
\end{tabular}
\end{table}
Table~\ref{tab:eer_asvspoof} summarizes the EER performance of several ASVspoof baseline systems in comparison with the proposed EDLFs-LR framework.
\begin{table}[ht]
\centering
\caption{EER Comparison with ASVspoof Baselines}
\label{tab:eer_asvspoof}
\begin{tabular}{lcc}
\hline
\textbf{Model} & \textbf{EER} & \textbf{Threshold} \\
\hline
LFCC-LCNN & 0.4192 & -5.84125 \\
GMM        & 0.4442 & -0.42018 \\
RawNet2     & 0.4442 & -1.63860 \\
EDLFs-LR          & 0.1611 & 0.30800 \\
\hline
\end{tabular}
\end{table}
\textbf{Repeated Holdout Validation}
To evaluate the robustness and stability of the proposed model, repeated random train-test split validation, also known as repeated holdout validation, was conducted. Initially, 50 repetitions were performed using random 80\%/20\% train-test splits. The stability of the evaluation metrics, including ROC AUC and EER, was monitored across repetitions. If the mean and standard deviation continued to vary significantly, the number of repetitions would be increased. After 50 repeated experiments, the following results were obtained for the best-performing model reported previously in Table \ref{tab:appendix_2} of the Appendix (LR using all EDLFs features):

\begin{itemize}
    \item Mean EER: 0.1734
    \item Standard Deviation of EER: 0.0310
    \item Mean ROC AUC: 0.8623
    \item Standard Deviation of ROC AUC: 0.0262
\end{itemize}

The corresponding confidence intervals are reported below:

\begin{itemize}
    \item EER: 0.1734 [0.1648, 0.1820]
    \item ROC AUC: 0.8623 [0.8551, 0.8696]
\end{itemize}

The relatively low standard deviation values indicate that the model performance remains stable across repeated experiments, demonstrating the robustness of the proposed framework.
\newline
\textbf{Significance Testing}

A Monte Carlo Simulation for significance testing was further conducted to assess whether the observed model performance could occur by chance. The testing procedure involved the following steps:

\begin{enumerate}
    \item The relationship between features and labels was disrupted by randomly shuffling the true labels.
    \item The model was retrained 1000 times using the randomized labels.
    \item Null distributions for both EER and ROC AUC were generated.
    \item The actual model performance was compared against the null distributions.
\end{enumerate}

The resulting p-values for both ROC AUC and EER were approximately 0.0001, indicating that the observed performance is highly unlikely under the null hypothesis of no meaningful relationship between the extracted features and the corresponding labels.

\subsubsection{EDLF Holdout Analysis}
\begin{figure}
 \includegraphics[width=.5\textwidth]{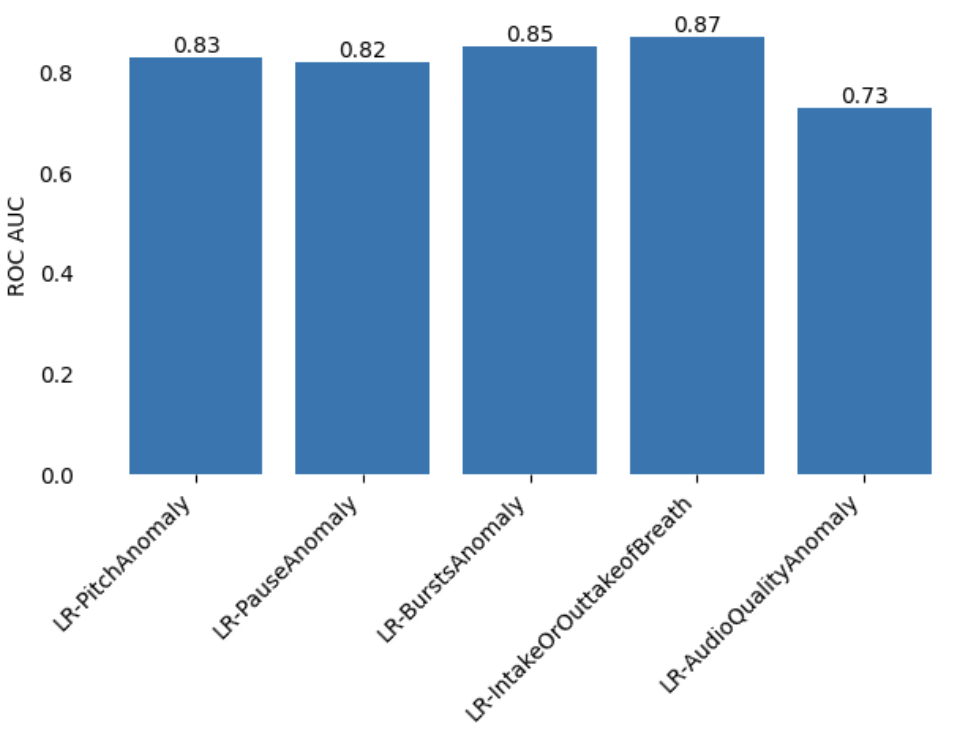}
  \caption{Dropout Analysis}
   \label{drop}
\end{figure}

Next, we conducted ablation studies where each EDLF was held out of the model to better understand the contribution of each EDLF to model performance. Here, we report the  holdout analysis for logistic regression only as shown in {\bf Figure \ref{drop}}. The holdout analysis was not significantly different across ML methods.
(Appendix A.1 Table \ref{tab:appendix_2} reports the LR model performance metrics for each held-out EDLF and the hyper parameters for LR are shown in Table \ref{tab:appendix_3}). 
We found that the audio quality EDLF is essential for optimal model performance. As illustrated in {\bf Figure \ref{drop}}, the AUC for the LR model dipped from 0.85 to 0.73 when all features except audio quality were present. Holding out each of the other four EDLFs had a negligible impact on LR prediction accuracy. Because we showed that  holding out individual EDLFs  influences the model outcome, this suggests that feature learning by the logistic regression model can be human-interpretable, which is an important consideration for explainability. A future study with a comprehensive ablation analysis can fully explore the impact of EDLF feature combinations on model performance.

\section{Discussion of Results}
We summarize several findings that support the linguistically augmented LinguAS dataset. 
\begin{itemize}
    \item We used LinguAS as training data for traditional ML models, finding that the best performing model on our multi attack dataset is logistic regression.   
    \item Our experimental results illustrated that any model augmented with the EDLF-LR model--ranging from traditional ML to SOTA SSL models--showed improved overall performance on unseen data. 
    \item In our LinguAS model testing, we find that incorporating multiple linguistic cues further improved model performance. We compared and augmented common ASVspoof baselines (LFCC-LCNN, LFCC-GMM, and Rawnet2). We also tested and augmented the SOTA models, including pre-trained HuBERT and XLSR representations input to MLP, as well as a CNN-based pre-trained model (VGGish-MLP).  Even considering the small size of the training data, LinguAS still improved all of the models. 
    \item LinguAS training data have flexibility from “deep data” rather than “large data.”  The narrow training data led to SOTA's lower performance; however, as indicated in the paper, LinguAS works even when the size of the training data is small. Further, LinguAS allows researchers to customize the  granularity of features they wish to use in deepfake detection models to use a combination of features, all features, or just one feature to evaluate model improvements. 
    \item Testing for which kind of models perform best with EDLF augmentation will be a useful direction to pursue. Such analysis will help indicate whether models are learning generator-specific cues or generalizable ones. LinguAS contains the generator/ source information for spoofed audio and can provide solid grounding for further research. 
    \item Our prior research shows that EDLFs are especially beneficial for certain kinds of voice spoofing attacks \cite{Khanjani2023}. Future research can investigate whether this finding also applies to the interaction between model architectures (deep learning, SSL) and attack types. Additionally, our prior work on causal cues in deepfake detection  \cite{Khanjani2024Investigating} investigated causal links of features with regard to deepfake detection, finding that anomalous audio quality was a front line feature. That being said, more research is needed in this area, as not all spoofed audio has audio quality issues, and not all genuine audio has high audio quality.
    \item Finally, the documentation in the LinguAS dataset clearly delineates how linguistic features should be classified in future development of linguistic feature-extracted preprocessing. We propose that the integration of expert linguists’ knowledge into the field of deepfake detection research improves communication across different research teams and clearer reporting of findings. 

    \end{itemize}

Our experiments validated the usefulness of EDLF-augmented data for improving deepfake detection algorithms and illustrated that they are useful data for training, fine-tuning, or testing model explainability. We contribute the LinguAS dataset to the research community to illustrate how focused, expert-annotated data can be more powerful for model inference than massive datasets.

\section{Ethical Considerations}
The audio speech data in LinguAS is entirely collected from publicly-available data sources, including open access datasets and public Youtube videos. We propose that including linguists in the development of linguistic feature representations improves ethical modeling and prevents algorithmic bias. As one example, linguists can ensure that diverse human speaker populations are proportionally represented in deepfake detection methods. In other areas of language technology development such as automatic speech recognition, systematic bias against minority speaker groups, such as speakers of African American English, have been identified \cite{Koenecke2020}. It is important to consider that a similar systematic bias could surface in deepfake detection algorithms, where certain dialects are poorly identified by the model as potentially deepfaked. 

\section{Limitations and Future Directions}
One limitation of the LinguAS dataset is that it contains English language samples only. This methodological choice is by design, because some linguistic features inherently differ in their generalizability across languages. For example, pitch is related to phonological meaning in Mandarin, while in English pitch variation is used for phrase-level and social meaning \cite{Duanmu2007, Jun2005}. Future work can include multiple languages tagged with metadata to allow for model comparison of what linguistic features are language-universal versus features that are language-specific.
Another challenge of the dataset creation for LinguAS is the annotation of features by expert linguists. The intent of the LinguAS data project is not to continually produce hand-annotated expert linguistic data. Instead, the LinguAS data is a pioneer project to demonstrate how existing knowledge of acoustic language variation from linguistics can be used to strategically identify features for dataset augmentation based on expert language scientist input. To improve generalizability, future versions of the LinguAS dataset will include algorithmically-calculated coefficient values for the linguistic features, thereby limiting the need for human input on each sample. Our preliminary work advances this direction through the autolabeling of EDLFs \cite{Khanjani2024ALDAS}.
In future work, we will expand our framework by identifying new, potentially informative linguistic features that are adaptively chosen by expert linguists based on their experience with differences between real human speech and fake speech. This incorporates an element of human-in-the-loop strategization into the development of deepfake detection models, which is currently lacking from research in this field. 

\section{Conclusion}
We introduced LinguAS, a dataset of over 800 speech audio samples evaluated and annotated by expert linguists on linguistic cues spanning multiple timescales that are indicative of real human speech. We produced a balanced dataset  of four attack types and genuine speech, also balanced by speaker gender across faked and genuine speech samples. The high-level, expert-annotated data provided in LinguAS can be used to create computationally-light and interpretable models based on a range of acoustic cues. In experiments that compared LinguAS-augmented data to baselines, we found significant improvements in model performance for every model we tested across model types and architectures. The LinguAS dataset is a timely contribution that can address current concerns in deepfake detection modeling techniques including computational cost and  generator-specific cue learning. The dataset is open access at \cite{figs}. 

\section{Acknowledgements}
Research reported in this publication was supported in part by the National Science Foundation, award numbers \#2210011 and \#2346473. Authors would like to thank Chloe Evered and Lavon Davis for their contributions to data collection, linguistic annotation, and data quality checks. All code is available at \cite{code} and dataset is available at \cite{figs}. The data description and data fame is available at \cite{dframe}.

\bibliographystyle{ACM-Reference-Format}
\bibliography{sample-base}

@article{Almutairi2022,
  author = {Almutairi, Z. and Elgibreen, H.},
  title = {{A review of modern audio deepfake detection methods: Challenges and future directions}},
  journal = {Algorithms},
  volume = {15},
  number = {5},
  pages = {155},
  year = {2022}
}

@article{Khochare2021,
  author = {Khochare, J. and Joshi, C. and Yenarkar, B. and Suratkar, S. and Kazi, F.},
  title = {{A deep learning framework for audio deepfake detection}},
  journal = {Arabian Journal for Science and Engineering},
  pages = {1--12},
  year = {2021}
}

@inproceedings{Sahidullah,
  author = {Sahidullah, M. and Kinnunen, T. and Hanil\c{C}i, C.},
  title = {{A comparison of features for synthetic speech detection}},
  note = {{Year and conference details are missing from the reference text.}},
  year = {}
}

@article{Iskarous2025,
  author = {Iskarous, K. and Vietti, A.},
  title = {{Phonetic information in the vowel spectrum: the meaning of mel-Frequency Cepstral Coefficients}},
  journal = {Journal of Phonetics},
  volume = {112},
  pages = {101434},
  year = {2025}
}

@inproceedings{Khanjani2023,
  author = {Khanjani, Z. and Davis, L. and Tuz, A. and Nwosu, K. and Mallinson, C. and Janeja, V. P.},
  title = {{Learning to Listen and Listening to Learn: Spoofed Audio Detection Through Linguistic Data Augmentation}},
  booktitle = {{2023 IEEE International Conference on Intelligence and Security Informatics (ISI)}},
  address = {Charlotte, NC, USA},
  pages = {01--06},
  year = {2023}
}

@article{Khanjani2024Investigating,
  author = {Khanjani, Z. and Ale, T. and Wang, J. and Davis, L. and Mallinson, C. and Janeja, V.},
  title = {{Investigating Causal Cues: Strengthening Spoofed Audio Detection with Human-Discernible Linguistic Features}},
  year = {2024},
  eprint = {2409.06033},
  archivePrefix = {arXiv},
  doi = {arXiv:2409.06033}
}

@inproceedings{Blue2022,
  author = {Blue, L. and Warren, K. and Abdullah, H. and Gibson, C. and Vargas, L. and O'Dell, J. and Butler, K. and Traynor, P.},
  title = {{Who are you (i really wanna know)? detecting audio DeepFakes through vocal tract reconstruction}},
  booktitle = {{31st USENIX Security Symposium (USENIX Security 22)}},
  pages = {2691--2708},
  year = {2022}
}

@inproceedings{Doan2023,
  author = {Doan, T. P. and Nguyen-Vu, L. and Jung, S. and Hong, K.},
  title = {{Bts-e: Audio deepfake detection using breathing-talking-silence encoder}},
  booktitle = {{ICASSP 2023-2023 IEEE International Conference on Acoustics, Speech and Signal Processing (ICASSP)}},
  pages = {1--5},
  year = {2023}
}

@inproceedings{Mostaani2022,
  author = {Mostaani, Z. and Doss, M.M.},
  title = {{On Breathing Pattern Information in Synthetic Speech}},
  booktitle = {{Proc. Interspeech 2022}},
  pages = {2768--2772},
  year = {2022},
  doi = {10.21437/Interspeech.2022-10271}
}

@inproceedings{Huang2019,
  author = {Huang, L. and Pun, C.-M.},
  title = {{Audio replay spoof attack detection using segment-based hybrid feature and DenseNetLSTM network}},
  booktitle = {{ICASSP 2019 - 2019 IEEE International Conference on Acoustics, Speech and Signal Processing (ICASSP)}},
  pages = {2567--2571},
  year = {2019}
}

@inproceedings{Kallay2019,
  author = {Kallay, J. E. and Mayr, U. and Redford, M. A.},
  title = {{Characterizing the coordination of speech production and breathing}},
  booktitle = {{Proceedings of the... International Congress of Phonetic Sciences}},
  volume = {2019},
  pages = {1412},
  year = {2019}
}

@article{RochetCapellan2014,
  author = {Rochet-Capellan, A. and Fuchs, S.},
  title = {{Take a breath and take the turn: how breathing meets turns in spontaneous dialogue}},
  journal = {{Philosophical Transactions of the Royal Society B: Biological Sciences}},
  volume = {369},
  number = {1658},
  year = {2014},
  note = {Pages: 20130399}
}

@inproceedings{Lai2019,
  author = {Lai, C.-I. and Abad, A. and Richmond, K. and Yamagishi, J. and Dehak, N. and King, S.},
  title = {{Attentive filtering networks for audio replay attack detection}},
  booktitle = {{ICASSP 2019 - 2019 IEEE International Conference on Acoustics, Speech and Signal Processing (ICASSP)}},
  pages = {6316--6320},
  year = {2019}
}

@article{Koenecke2020,
  author = {Koenecke, A. and Nam, A. and Lake, E. and Nudell, J. and Quartey, M. and Mengesha, Z. and others and Goel, S.},
  title = {{Racial disparities in automated speech recognition}},
  journal = {{Proceedings of the national academy of sciences}},
  volume = {117},
  number = {14},
  pages = {7684--7689},
  year = {2020}
}

@book{Duanmu2007,
  author = {Duanmu, S.},
  title = {{The phonology of standard Chinese}},
  publisher = {{Oxford University Press}},
  year = {2007}
}

@book{Jun2005,
  author = {Jun, S. A.},
  title = {{Prosodic typology: the phonology of intonation and phrasing}},
  publisher = {{Oxford University Press}},
  year = {2005}
}

@article{Akhtar2024,
  author = {Akhtar, Z. and Pendyala, T. L. and Athmakuri, V. S.},
  title = {{Video and Audio Deepfake Datasets and Open Issues in Deepfake Technology: Being Ahead of the Curve}},
  journal = {Forensic Sciences},
  volume = {4},
  number = {3},
  pages = {289--377},
  year = {2024},
  doi = {10.3390/forensicsci4030021}
}

@article{Chodroff2014,
  author = {Chodroff, E. and Wilson, C.},
  title = {{Burst spectrum as a cue for the stop voicing contrast in American English}},
  journal = {The Journal of the Acoustical Society of America},
  volume = {136},
  number = {5},
  pages = {2762--2772},
  year = {2014},
  doi = {10.1121/1.4896470}
}

@article{Hong2025,
  author = {Hong, J. and Chung, Y. and Oh, S. and Kim, J. and Lee, J. and Kim, S. and Cho, H.},
  title = {{TwinShift: Benchmarking Audio Deepfake Detection across Synthesizer and Speaker Shifts}},
  journal = {arXiv preprint arXiv:2510.23096},
  year = {2025},
  eprint = {2510.23096},
  archiveprefix = {arXiv},
  primaryClass = {cs.SD}
}

@article{Khanjani2024ALDAS,
  author = {Khanjani, Z. and Mallinson, C. and Foulds, J. and Janeja, V. P.},
  title = {{ALDAS: Audio-Linguistic Data Augmentation for Spoofed Audio Detection}},
  journal = {arXiv preprint arXiv:2410.15577},
  year = {2024},
  eprint = {2410.15577},
  archiveprefix = {arXiv},
  primaryClass = {cs.SD}
}

@inproceedings{MartinDonas2024,
  author = {Mart{\'{i}}n-Do{\~{n}}as, J. M. and {\'{A}}lvarez, A. and Rosello, E. and Gomez, A. M. and Peinado, A. M.},
  title = {{Exploring Self-supervised Embeddings and Synthetic Data Augmentation for Robust Audio Deepfake Detection}},
  booktitle = {{Interspeech 2024}},
  pages = {2085--2089},
  year = {2024},
  doi = {10.21437/Interspeech.2024-942}
}

@article{Pindrop2025,
  author = {{Pindrop}},
  title = {{Pindrop’s 2025 Voice Intelligence and Security Report Reveals 1,300\% Surge in Deepfake Fraud}},
  journal = {{PR Newswire}},
  month = {June},
  day = {12},
  year = {2025},
  url = {https://www.prnewswire.com/news-releases/pindrops-2025-voice-intelligence--security-report-reveals-1-300-surge-in-deepfake-fraud-302479482.html}
}

@article{Zhang2025,
  author = {Zhang, K. and Hua, Z. and Lan, R. and Zhang, Y. and Guo, Y.},
  title = {{Phoneme-Level Feature Discrepancies: A Key to Detecting Sophisticated Speech Deepfakes}},
  journal = {{Proceedings of the AAAI Conference on Artificial Intelligence}},
  volume = {39},
  number = {1},
  pages = {1066--1074},
  year = {2025},
  doi = {10.1609/aaai.v39i1.32093}
}

@article{Park2020Cotatron,
  author = {Park, S. W. and Kim, D. Y. and Joe, M. C.},
  title = {{Cotatron: Transcription-guided speech encoder for any-to-many voice conversion without parallel data}},
  journal = {arXiv preprint arXiv:2005.03295},
  year = {2020},
  eprint = {2005.03295},
  archivePrefix = {arXiv},
  primaryClass = {eess.AS}
}

@inproceedings{Valle2020Mellotron,
  author = {Valle, R. and Li, J. and Prenger, R. and Catanzaro, B.},
  title = {{Mellotron: Multispeaker expressive voice synthesis by conditioning on rhythm, pitch and global style tokens}},
  booktitle = {{ICASSP 2020 - 2020 IEEE International Conference on Acoustics, Speech and Signal Processing (ICASSP)}},
  pages = {6189--6193},
  year = {2020},
  organization = {IEEE}
}

@article{VanDenOord2016Wavenet,
  author = {Van Den Oord, A. and Dieleman, S. and Zen, H. and Simonyan, K. and Vinyals, O. and Graves, A. and others and Kavukcuoglu, K.},
  title = {{Wavenet: A generative model for raw audio}},
  journal = {arXiv preprint arXiv:1609.03499},
  volume = {12},
  number = {1},
  year = {2016},
  eprint = {1609.03499},
  archivePrefix = {arXiv},
  primaryClass = {cs.SD}
}

@article{Ito2017LJSpeech,
  author = {Ito, K. and Johnson, L.},
  title = {{The LJ speech dataset}},
  year = {2017}
}

@inproceedings{Kim2022AssemVC,
  author = {Kim, K.-W. and Park, S.-W. and Lee, J. and Joe, M.-C.},
  title = {{Assem-vc: Realistic voice conversion by assembling modern speech synthesis techniques}},
  booktitle = {{ICASSP 2022 - 2022 IEEE International Conference on Acoustics, Speech and Signal Processing (ICASSP)}},
  pages = {6997--7001},
  year = {2022}
}

@inproceedings{Kumar2019Melgan,
  author = {Kumar, K. and Kumar, R. and de Boissi{\`e}re, T. and Gestin, L. and Teoh, W. Z. and Sotelo, J. and de Br\'{e}bisson, A. and Bengio, Y. and Courville, A. C.},
  title = {{Melgan: Generative adversarial networks for conditional waveform synthesis}},
  booktitle = {{Advances in neural information processing systems 32}},
  year = {2019}
}

@inproceedings{Reimao2019FoR,
  author = {Reimao, R. and Tzerpos, V.},
  title = {{FoR: A dataset for synthetic speech detection}},
  booktitle = {{2019 International Conference on Speech Technology and Human-Computer Dialogue (SpeD)}},
  pages = {1--10},
  year = {2019}
}

@article{Wang2021Comparative,
  author = {Wang, X. and Yamagishi, J.},
  title = {{A comparative study on recent neural spoofing countermeasures for synthetic speech detection}},
  journal = {arXiv preprint arXiv:2103.11326},
  year = {2021},
  eprint = {2103.11326},
  archivePrefix = {arXiv},
  primaryClass = {cs.SD}
}

@article{Wu2017ASVspoof,
  author = {Wu, Z. and Yamagishi, J. and Kinnunen, T. and Hanil\c{C}i, C. and Sahidullah, M. and Sizov, A. and Evans, N. and Todisco, M. and Delgado, H.},
  title = {{Asvspoof: the automatic speaker verification spoofing and countermeasures challenge}},
  journal = {{IEEE Journal of Selected Topics in Signal Processing}},
  volume = {11},
  number = {4},
  pages = {588--604},
  year = {2017}
}

@inproceedings{Kinnunen2017ASVspoof,
  author = {Kinnunen, T. and Sahidullah, M. and Delgado, H. and Todisco, M. and Evans, N. and Yamagishi, J. and Lee, K. A.},
  title = {{The ASVspoof 2017 challenge: Assessing the limits of replay spoofing attack detection}},
  booktitle = {{Interspeech 2017}},
  pages = {26},
  year = {2017},
  organization = {ISCA}
}

@article{Yamagishi2021ASVspoof2021,
  author = {Yamagishi, J. and Wang, X. and Todisco, M. and Sahidullah, M. and Patino, J. and Nautsch, A. and Liu, X. and Lee, K. A. and Kinnunen, T. and Evans, N. and others},
  title = {{Asvspoof 2021: accelerating progress in spoofed and deepfake speech detection}},
  journal = {arXiv preprint arXiv:2109.00537},
  year = {2021},
  eprint = {2109.00537},
  archivePrefix = {arXiv},
  primaryClass = {cs.SD}
}

@article{figs,
  author = {Keaton, Ashley R. and Khanjani, Zahra and Mallinson, Christine and Janeja, Vandana P.},
  title = {{Linguistically Augmented Audio Speech Data (LinguAS) Dataset}},
  howpublished = {\url{ https://figshare.com/projects/_b_Linguistically_Augmented_Audio_Speech_Data_LinguAS_b_/206566}},
  note = {Data available at Figshare},
  year = {2024} 
}

@article{code,
  author = {Khanjani, Zahra and Keaton, Ashley R. and Mallinson, Christine and Janeja, Vandana P.},
  title = {{LinguAS Code Repository}},
  howpublished = {\url{https://github.com/MultiDataLab/LinguAS}},
  note = {Source code available at GitHub},
  year = {2024}
}

@inproceedings{dframe,
  author    = {Ashley R. Keaton and Zahra Khanjani and Christine Mallinson and Vandana P. Janeja},
  title     = {Linguistically Augmented Audio Speech Data (LinguAS), Direct link for Data frame and description},
  year      = {2026},
  url       = {https://doi.org/10.6084/m9.figshare.25909297}
}

@article{Strickland,
  author  = {Strickland, Eliza},
  title   = {Andrew Ng, AI Minimalist: The Machine-Learning Pioneer Says Small Is the New Big},
  journal = {IEEE Spectrum},
  year    = {2022},
  volume  = {59},
  number  = {4},
  pages   = {22--50}
}

@article{WestburyKeating1986,
  author  = {J. R. Westbury and P. A. Keating},
  title   = {On the Naturalness of Stop Consonant Voicing},
  journal = {Journal of Linguistics},
  volume  = {22},
  number  = {1},
  pages   = {145--166},
  year    = {1986},
  doi     = {10.1017/s0022226700010598}
}

@article{muller,
  title={Does audio deepfake detection generalize?},
  author={M{\"u}ller, Nicolas M and Czempin, Pavel and Dieckmann, Franziska and Froghyar, Adam and B{\"o}ttinger, Konstantin},
  journal={arXiv preprint arXiv:2203.16263},
  year={2022}
}
\section*{Appendix}

\subsection*{A.1 Dataset Metadata}

Appendix

A.1 Dataset Metadata
\begin{table}[htbp]
\centering
\setlength{\tabcolsep}{3pt} 
\renewcommand{\arraystretch}{1.15} 
\caption{Distribution of Fake Audio Attack Type}
\label{tab:appendix_1}

\begin{tabular}{l c c c c} 
\hline
\textbf{Type} & \textbf{Source} & \textbf{Qty.} & \textbf{Male Prop.} & \textbf{Spoofed Prop.} \\
\hline

Replay Attack & ASVspoof 2017 & 103 & 0.70 & 0.22 \\
TTS & ASVspoof 2021 & 33 & 0.48 & 0.07 \\
TTS & Google-TTS (Generated by LinguAS research team) & 18 & 0.55 & 0.03 \\
TTS & MelGAN (Generated by LinguAS research team) & 15 & 0.13 & 0.03 \\
TTS & WaveNet & 19 & 0.63 & 0.04 \\
TTS & Descript & 14 & 1.00 & 0.03 \\
TTS & FoR dataset & 35 & 0.48 & 0.07 \\
VC & ASVspoof 2021 & 88 & 0.60 & 0.18 \\
VC & Assem-VC & 29 & 0.27 & 0.06 \\
VC & resemble.ai (Generated by LinguAS research team) & 8 & 0.75 & 0.01 \\
VC & PPG & 8 & 0.62 & 0.01 \\
VC & Mellotron & 7 & 0.57 & 0.01 \\
VC & Cotatron & 7 & 0.57 & 0.01 \\
Mimicry & YouTube & 82 & 0.64 & 0.175 \\
\hline
\textbf{Total} & & \textbf{466} & \textbf{0.59} & \textbf{1.00} \\
\hline
\end{tabular}
\end{table}


\begin{table}[h!]
\caption{Comparison of Logistic Regression Model Performance for each held-out EDLF}
\centering
\label{tab:appendix_2}
\begin{tabular}{l | l | l | l | l | l}
\hline
Held-out feature & False positive rate & True positive rate & Accuracy & AUC score \\
\hline
Audio Quality Anomaly & 0.123 & 0.539 & 0.684 & 0.732 \\
\hline
Breath Anomaly & 0.070 & 0.697 & 0.797 & 0.866 \\
\hline
Pitch Anomaly & 0.070 & 0.697 & 0.797 & 0.835 \\
\hline
Bursts Anomaly & 0.070 & 0.697 & 0.797 & 0.847 \\
\hline
Pause Anomaly & 0.070 & 0.697 & 0.797 & 0.825 \\
\hline
LR with all EDLF features & 0.070 & 0.697 & 0.797 & 0.850 \\
\hline
\end{tabular}
\end{table}

\begin{table}[h!]
\caption{Logistic Regression Hyper Parameters}
\centering
\label{tab:appendix_3}
\begin{tabular}{l | l | l | l | l | l}
\hline
Held-out feature & Penalty & C & Solver & Max iter \\
\hline
Audio Quality Anomaly & l2 & 0.1 & lbfgs & 1000 \\
\hline
Breath Anomaly & l2 & 0.1 & lbfgs & 1000 \\
\hline
Pitch Anomaly & l2 & 0.1 & lbfgs & 1000 \\
\hline
Bursts Anomaly & l2 & 0.1 & lbfgs & 1000 \\
\hline
Pause Anomaly & l2 & 10 & lbfgs & 1000 \\
\hline
LR with all EDLF features & l2 & 0.1 & lbfgs & 1000 \\
\hline
\end{tabular}
\end{table}
\end{document}